# Accurate color imaging of pathology slides using holography and absorbance spectrum estimation of histochemical stains

*Yibo Zhang[1,2,3,†], Tairan Liu[1,2,3,†], Yujia Huang[1], Da Teng[4], Yinxu Bian[1], Yichen Wu[1,2,3], Yair Rivenson[1,2,3], Alborz Feizi[1,2,3,5], and Aydogan Ozcan[1,2,3,6,*]*

[†] Equally contributing authors

*Corresponding author: E-mail: ozcan@ucla.edu

[1]Electrical and Computer Engineering Department, University of California, Los Angeles, CA, 90095, USA.
[2]Bioengineering Department, University of California, Los Angeles, CA, 90095, USA.
[3]California NanoSystems Institute (CNSI), University of California, Los Angeles, CA, 90095, USA.
[4]Computer Science Department, University of California, Los Angeles, CA, 90095, USA.
[5]Department of Biomedical Engineering, Yale University, New Haven, CT 06520, USA
[6]Department of Surgery, David Geffen School of Medicine, University of California, Los Angeles, CA, 90095, USA.

Keywords: digital holography, computational imaging, absorbance spectrum estimation, color holography

**Abstract**: Holographic microscopy presents challenges for color reproduction due to the usage of narrow-band illumination sources, which especially impacts the imaging of stained pathology slides for clinical diagnoses. Here, an accurate color holographic microscopy framework using absorbance spectrum estimation is presented. This method uses multispectral holographic images acquired and reconstructed at a small number (e.g., three to six) of wavelengths, estimates the absorbance spectrum of the sample, and projects it onto a color tristimulus. Using this method, the wavelength selection is optimized to holographically image 25 pathology slide samples with different tissue and stain combinations to significantly reduce color errors in the final reconstructed images. The results can be used as a practical guide for various imaging applications and, in particular, to correct color distortions in holographic imaging of pathology samples spanning different dyes and tissue types.

## 1. Introduction

Digital holographic microcopy has been widely applied in various biomedical-imaging tasks owing to its unique advantages such as extended depth-of-field, simple optical configuration, and the capability to create amplitude and phase images of the sample [1–26]. Recently, digital holography using a lens-free on-chip configuration has been employed to image pathology samples including thin-sectioned tissues, cell smears, and even optically cleared thick tissues in three dimensions (3D) for cell phenotyping and disease diagnosis. This method allows to build cost-effective, field-portable, and wide field-of-view (FOV) microscopes that can be deployed in low-resource regions. Based on phase retrieval [4,7,11,27,28] and resolution enhancement [2,4,5,9–11,29–33] techniques, wide-field and high-resolution images are rapidly generated through computation.

Generation of images with accurate color reproduction is one of the key requirements for a digital pathology microscope to ensure reliable and consistent diagnoses [34–37]. Among other approaches, color images can be generated by a holographic microscope through the combination of three holograms acquired at three discrete wavelengths in the red, green, and blue regions of the visible light spectrum. The three intensity images are then placed into the R, G, and B channels of an RGB image ("direct RGB combination" method). In other words, the spectral transmittance or reflectance of the sample is *discretely* sampled at three chosen wavelengths and directly used to compose a color image. However, this approach suffers from color inaccuracies [38] compared to the color perception of the human eye, which has broad responsivity curves according to the Commission Internationale de l'Éclairage (CIE) color matching functions [39].

Prior research on improving the color accuracy of holographic imaging has mainly focused on imaging generic objects and scenes in the reflection mode. For example, Peercy *et al.* improved the accuracy of color holography for the reconstruction of macroscopic scenes based on the Gaussian quadrature and Riemann summation methods, and provided an optimal wavelength selection, showing that four wavelengths could lead to a relatively accurate color reproduction [38]. Later, Ito *et al.* presented a spectral-estimation approach based on the Wiener estimation for accurate-color holography using laser illuminations at four wavelengths [40]. While these approaches have been applied to generic objects in the reflection mode, accurate-color holographic imaging of stained pathology samples in the transmission mode remains largely unexplored. Compared to the colors of generic objects that cover a large subspace of all possible colors, the colors that occur within a stained pathology slide are by and large constrained by the dye combination that is used to stain the slide. The combination typically consists of two or three dyes and is usually known *a priori*, before the slide is imaged [41]. Consequently, it is important to develop a method to achieve accurate-color holographic imaging of pathology slides with a smaller number of wavelengths as well as better color accuracy.

In this paper, we present an accurate-color holographic imaging method for pathology slide imaging using an absorbance spectrum estimation-based colorization (ASEC) technique. We demonstrate this technique using a lens-free on-chip holographic microscope (**Figure 1**), but the method is broadly applicable to other holographic imaging systems. The imaging setup (Figure 1(a)) uses a partially coherent source that emits narrow-band light centered at a single wavelength or multiple wavelengths *simultaneously* to illuminate the sample (e.g., a pathology slide). Further, the setup employs a complementary metal-oxide-semiconductor (CMOS) image sensor to capture the diffraction patterns (holograms) of the sample. The image sensor setup comprises a 3D translation stage to capture multiple holograms, which are processed with the subsequent pixel super-resolution (PSR) [2,5,9,10,29–33] and phase recovery [2,5,7,27,28] algorithms to generate a high-resolution image of the sample under multiple illumination wavelengths. The images are then fed to the ASEC algorithm that has been pretrained on various tissue types and stains. The ASEC algorithm uses *a priori* information about the spectral statistics of each given tissue type and stain combination to estimate the absorbance spectrum of the sample at each pixel, yielding an estimated hyperspectral cube of the sample. This hyperspectral cube is then used to simulate the color response of the human eye to create a color-accurate holographic image [39].

Based on the ASEC framework, we performed wavelength optimizations for 25 pathology slides with different tissue–stain combinations using various numbers of wavelengths ($n = 3, 4, 5,$ or 6), and tested the wavelength-optimized ASEC method on all 25 samples. The results show that using three illumination wavelengths ($n = 3$) is sufficient to create color-accurate reconstruction images of the samples. More wavelengths ($n = 4, 5,$ or 6) further reduce color errors. In addition, it is demonstrated that the ASEC method can be applied in conjunction with wavelength-multiplexed holographic imaging using the demosaiced pixel super-resolution (DPSR) technique, which reduces the data acquisition time three-fold.

## 2. Materials and Methods
### 2.1. Lens-free imaging setup

As shown in Figure 1(a), a broad-band light source (WhiteLase Micro, NKT Photonics) filtered by an acousto-optic tunable filter (AOTF) with a spectral bandwidth of approximately 2.5 nm (full width at half maximum, i.e., FWHM) was used as illumination source. The AOTF supports simultaneously up to eight output channels at different wavelengths. All wavelength channels can be coupled into the same single-mode fiber. We directly used the emitted, uncollimated light from the single-mode fiber to create quasi-plane-wave illumination for the sample (e.g., a pathology slide) with a light source-to-sample distance ($z_1$) of approximately 5–10 cm. A CMOS image sensor chip (IMX 081, Sony, pixel size of 1.12 μm) was placed below the sample with a small sample-to-sensor distance ($z_2$), which is typically < 2 mm. The CMOS image sensor was attached to a 3D positioning stage (MAX606, Thorlabs, Inc.). The translation was used for PSR [2,4,7,9,32,33], DPSR [10], and multi-height phase recovery techniques [2,5,7,12,32,33]. All the hardware was automatically controlled by LabVIEW.

### 2.2. Multispectral lens-free imaging at three wavelengths using wavelength multiplexing

The tunable light source was programmed to *simultaneously* output narrow-band light at three wavelengths, corresponding to the red, green, and blue regions of the visible spectrum [10]. The mechanical stage was programmed to shift the image sensor on a lateral 6 × 6 grid (step size of approximately 0.37 μm, corresponding to 1/3 of the pixel size) at each hologram acquisition height for DPSR and to $N_z$ different heights for the multi-height phase recovery, where one "composite hologram" that results from the simultaneous three-wavelength illumination was captured by the image sensor at each of the 6 × 6 × $N_z$ recording positions. The acquired data were then processed using DPSR, multi-height phase recovery, and ASEC, which will be detailed in the following subsections.

### 2.3. DPSR technique

While the color CMOS image sensor was being laterally shifted on a 6 × 6 grid with sub-pixel distances, a hologram was recorded at each position. These shifts do not need to be accurate because the actual amounts of shifting were precisely estimated using a shift estimation algorithm [2,4,5,9,10,32,33,42]. A shift-and-add based PSR method [7] was then used to synthesize four high-resolution pixel-super-resolved holograms; one for each color pixel in the Bayer pattern (R, $G_1$, $G_2$, and B) with a resulting effective pixel size of approximately 0.37 μm [7]. Each of these four synthesized high-resolution holograms is a combined response resulting from the transmitted and diffracted light intensity at multiple wavelengths that was multiplied with the spectral responsivity of the corresponding pixel (R, $G_1$, $G_2$, or B). Demosaicing (i.e., demultiplexing) was conducted to remove the wavelength crosstalk by multiplying with the pseudoinverse of the "crosstalk matrix" [10]. The crosstalk matrix can be estimated from the three chosen wavelengths and previously measured spectral responsivities of the four-pixel types in the Bayer pattern; the two green pixels in the Bayer pattern, $G_1$ and $G_2$, have slightly different responsivities, as experimentally measured. The demosaicing process results in three high-resolution holograms without spectral crosstalk. The DPSR algorithm was accelerated using a graphics processing unit (GPU) and CUDA C++ programming.

### 2.4. Computational back-propagation of holograms using the angular-spectrum method

We applied the angular-spectrum method, which is an essential building block in the multi-height phase recovery and holographic autofocusing, to computationally back-propagate holograms and complex optical fields [43]. The complex-valued optical field at $z = z_0$, $U(x,y; z_0)$, was first transformed to the angular spectrum (spatial frequency) domain via a fast Fourier transform (FFT). The angular spectrum at $z = z_0$ was then multiplied with a spatial frequency-dependent phase factor parametrized by the wavelength, refractive index of the medium, and propagation distance ($\Delta z$) [43,44] to obtain the angular spectrum at $z = z_0 + \Delta z$. Finally, the optical field was back-transformed to the spatial domain via an inverse FFT, resulting in $U(x,y; z_0+\Delta z)$. The complex optical field $U(x,y; z)$ is related to the light intensity $I(x,y; z)$ via $I(x,y; z) = |U(x,y; z)|^2$.

### 2.5. Transport-of-intensity equation (TIE) and multi-height phase recovery

Holograms at eight heights were captured at approximate sample-to-sensor distances of 0, 15, 30, 45, 60, 75, 90, and 180 μm relative to the first height by moving the image sensor downwards (away from the sample) using the positioning stage. The z-shift amount do not need to be precise as the accurate z distances were estimated using a holographic autofocusing algorithm based on the edge sparsity criterion ("Tamura of the gradient", i.e., ToG) [45,46], detailed in 2.6. An initial estimation of the phase was obtained by applying TIE [7,47] to the (demosaiced) pixel-super-resolved holograms at the 1st, 7th, and 8th heights, which were approximately equally spaced. Then, an iterative multi-height phase recovery algorithm was used to further refine the phase: the current estimate of the complex optical wave was digitally propagated to each hologram height, where the amplitude was updated by averaging with the square-root of the pixel-super-resolved hologram intensity and the phase was kept unchanged [2,5,7,10,12,32,33]. The described update was executed starting from the farthest (8th) hologram height with respect to the image sensor back towards the nearest (1st) height, and directly to the $8^{th}$ height at the end, which defines one iteration. Within 10–30 iterations, the phase typically converges, and the amplitude is refined. The final step comprises back-propagating the complex wave defined by the converged amplitude and phase to the sample plane based on the distance found by the autofocusing algorithm [45,46]. The TIE solver was implemented through the elliptic equation solver in MATLAB [7,48,49]. The iterative multi-height phase recovery algorithm was accelerated using GPU and CUDA C++ programming.

### 2.6. Hologram autofocusing using the edge sparsity criterion

The z-distance of a given hologram is automatically estimated through the maximization of the ToG focus criterion proposed in [45,46] as a function of the back-propagation distance. To accelerate the maximization process, we custom-wrote a search algorithm that consists of a rough scan, followed by the determination of a unimodal interval in the vicinity of the maximum and a golden-section search within the unimodal interval. More implementation details can be found in [45]. The autofocusing algorithm was accelerated using GPU and CUDA C++ programming.

### 2.7. Multispectral lens-free imaging through sequential wavelength scanning

The aforementioned wavelength-multiplexed lens-free imaging method (Section 2.2) is highly time- and data-efficient but mostly limited to three (or multiples of three) wavelengths that belong to the red, green, and blue regions of the visible spectrum, respectively [10]. An alternative approach to perform multispectral (and hyperspectral) lens-free imaging at arbitrary wavelengths comprises the use of *sequential* multi-wavelength illumination, as opposed to simultaneous illumination at multiple wavelengths. At each scanning position during image acquisition for PSR and multi-height phase recovery, the wavelength-tunable illumination source was programmed to sequentially illuminate the sample at $n$ wavelengths (one wavelength at a given time). The image at each wavelength was individually captured by the image sensor. The PSR and multi-height phase recovery algorithms were then applied to the data collected under each wavelength, respectively, to reconstruct the sample images at the regarding wavelength. As a special case, while performing hyperspectral imaging, the sequentially scanned illumination wavelength range spanned from 400 nm to 700 nm with a step size of 10 nm.

### 2.8. ASEC for holographic reconstruction of stained tissues and cells

In general, for a transmission-mode holographic imaging system that uses a single wavelength, the *reconstructed* light intensity $I(x, y)$ at the sample plane can be given by [50]

$$I(x, y) = \int S(\lambda) T(x, y; \lambda) R(\lambda) d\lambda + \text{noise} \quad (1)$$

where $\lambda$ is the optical wavelength, $S(\lambda)$ the spectral intensity distribution of the illuminant, $T(x,y; \lambda)$ the spectral transmittance of the object, and $R(\lambda)$ the spectral responsivity of a given type of image sensor pixel (R, $G_1$, $G_2$, or B). When the spectrum of the illumination source is relatively narrow (centered around $\lambda_0$) for a typical holographic system, while $T(x,y; \lambda)$ and $R(\lambda)$ are slowly varying functions of $\lambda$ compared to the illumination light source bandwidth, Eq. (1) can be approximated as:

$$I(x, y) = S_0 T(x, y; \lambda_0) R(\lambda_0) + \text{noise} \quad (2)$$

where $S_0$ is the light intensity of the illuminant,

$$S_0 = \int S(\lambda) d\lambda \quad (3)$$

The background, where no sample is present, has a transmittance equal to unity. Therefore, assuming noise is negligible, the background-normalized intensity $\tilde{I}(x,y)$ of a sample can be written as:

$$\tilde{I}(x,y) = \frac{I(x,y)}{I_{background}} \approx \frac{S_0 R(\lambda_0) T(x,y;\lambda_0)}{S_0 R(\lambda_0)} = T(x,y;\lambda_0), \quad (4)$$

which is equal to the transmittance of the sample at $\lambda_0$ and independent of the illumination source intensity and pixel responsivity.

For our wavelength-multiplexed holographic imaging system, we have $n$ reconstructed images for $n$ wavelengths. The reconstructed intensity at each wavelength can then be normalized on the sample plane by the empty regions to obtain the transmittance function, $T(x, y; \lambda_i)$ for $i = 1, 2, \ldots, n$. The absorbance can be calculated from the transmittance function using:

$$A(x,y;\lambda_i) = -\log_{10}\left[T(x,y;\lambda_i)\right] \quad (5)$$

To estimate the entire absorbance spectrum, the minimum mean-square error (MMSE) method based on the Wiener estimation was used [50,51]. Suppose we want to reconstruct the absorbance spectrum discretized at $m$ wavelengths (e.g., $m = 31$ for spectral sample points at 400 nm, 410 nm, …, 700 nm), and we have available measurement data at $n$ wavelengths (e.g., $n = 3, 4, 5,$ or 6). The forward model, i.e., the spectral sampling operation, can be written as

$$\mathbf{g} = \mathbf{Ha} \quad (6)$$

where $\mathbf{g}$ is a $n \times 1$ vector representing the measured absorbance values at $n$ different wavelengths ($\lambda_i$, i = 1, 2, …, $n$); $\mathbf{a}$ is the $m \times 1$ absorbance spectrum that we want to recover discretized at $\lambda_j'$, $j = 1, 2, \ldots, m$; $\mathbf{H}$ is an $n \times m$ "sampling matrix"—the elements of which are given by

$$h_{ij} = \begin{cases} 1 & \text{if } \lambda_i = \lambda_j' \\ 0 & \text{else} \end{cases} \quad (7)$$

The solution to the MMSE problem is given as [51]:

$$\hat{\mathbf{a}} = \mathbf{W}\mathbf{g} \qquad (8)$$

$$\mathbf{W} = \mathbf{R}_{aa}\mathbf{H}^{T}\left(\mathbf{H}\mathbf{R}_{aa}\mathbf{H}^{T}\right)^{-1} \qquad (9)$$

where $\mathbf{R}_{aa}$ denotes the autocorrelation matrix of $\mathbf{a}$. Refer to section 2.10.2 and Eq. (12) for the process that is used to estimate $\mathbf{R}_{aa}$. The estimated transmittance spectrum $\mathbf{t} = [t_1, t_2, \ldots, t_m]^{T}$ can then be written as:

$$t_j = 10^{-\hat{a}_j} \qquad (10)$$

where $\hat{a}_j$ is the $j$-th element of $\hat{\mathbf{a}}$ with $j = 1, 2, \ldots, m$. The color tristimulus can then be calculated from the estimated spectrum (see Section 2.9).

## 2.9. Calculation of the color tristimulus from the transmittance spectrum of the sample

The tristimulus values, i.e., XYZ values [39], can be calculated from the transmittance spectrum of the sample using:

$$\begin{aligned} X &= \int \bar{x}(\lambda) T(\lambda) E(\lambda) \mathrm{d}\lambda \\ Y &= \int \bar{y}(\lambda) T(\lambda) E(\lambda) \mathrm{d}\lambda \\ Z &= \int \bar{z}(\lambda) T(\lambda) E(\lambda) \mathrm{d}\lambda \end{aligned} \qquad (11)$$

where $\bar{x}(\lambda)$, $\bar{y}(\lambda)$, and $\bar{z}(\lambda)$ are the CIE color matching functions [39]. They represent the spectral responsivities of the cone cells of an average human eye; $E(\lambda)$ is a standard illuminant, where the CIE Standard Illuminant D65 was used in this study [39]. The resulting XYZ tristimulus values can be converted to standard RGB values using a simple linear transformation [39].

## 2.10. Optimization of sample-specific illumination wavelengths

We developed a procedure (see **Figure 2**) to optimize the wavelength selection for each specific tissue–stain type. The details are presented below.

*2.10.1. Acquiring the ground truth*

In this study, we considered the color images that are calculated from the hyperspectral lens-free reconstructed images as the "ground truth" (left column in Figure 2). It was used for the wavelength optimization and evaluation of the ASEC method. Therefore, the hyperspectral lens-free imaging and reconstruction were conducted as detailed in Section 2.7. The reconstructed hyperspectral cube was converted to the color tristimulus (CIE XYZ color space; see Section 2.9). Three 2048-by-2048-pixel regions of interest (ROIs) that represent diverse spatial features of the sample were manually picked across the entire imaging FOV and were used for training. Details related to training and testing are presented below.

*2.10.2. Collection of spectral statistics and optimization of wavelength combinations (training)*

As shown in the middle column of Figure 2, we converted the hyperspectral transmittance images within the three training ROIs (2048-by-2048 pixels in each) to the hyperspectral absorbance by normalizing the intensity images using their respective average background values and applying $-\log_{10}()$ (see Eqs. (4) and (5)). All the pixels within the three training ROIs (background areas excluded) were used to approximate the autocorrelation matrix

$$\mathbf{R}_{\mathbf{aa}} \approx \frac{1}{N_{\text{pixels}}} \sum_{i \in P} \mathbf{a}_i \mathbf{a}_i^{\text{T}} \qquad (12)$$

where $N_{\text{pixels}}$ is the total number of pixels for averaging, and P is the set of all $N_{\text{pixels}}$ used for the $\mathbf{R}_{\mathbf{aa}}$ approximation. Once $\mathbf{R}_{\mathbf{aa}}$ is obtained for a given sample type (i.e., tissue–stain combination), it can be used to calculate $\mathbf{W}$ for any given wavelength combination for the same sample type (see Eq. (9)). Thus, ASEC can be readily performed when $\mathbf{R}_{\mathbf{aa}}$ is estimated.

For each combination of tissue type and stain, and given number of wavelengths ($n$), the optimal wavelength combination, after performing ASEC, leads to the minimum $\Delta E^*_{94}$ when compared to the ground truth color images. For each sample type, this wavelength optimization needs to be conducted only once. For this purpose, a total of 3,000 pixels were randomly chosen from the three training ROIs to be used in the wavelength optimization. An exhaustive search was conducted to obtain the globally optimal wavelength combination; i.e., for discrete wavelengths starting from 400 nm to 700 nm with 10 nm increment (31 in total), all $\binom{31}{n}$ wavelength combinations were calculated for a given $n$. For each of these $n$-wavelength combinations, the transmittance values corresponding to the randomly chosen 3,000 pixels were collected. Afterwards, the ASEC method was applied to each of these pixels to estimate their tristimulus values (CIE XYZ color space) using the previously calculated $\mathbf{R_{aa}}$ (Eqs. (8)–(12)). The average $\Delta E^*_{94}$ was determined considering the estimated tristimulus and ground truth tristimulus values. After calculating $\Delta E^*_{94}$ for all different wavelength combinations, the $\Delta E^*_{94}$ values were sorted, and the combination corresponding to the smallest $\Delta E^*_{94}$ was selected as optimal combination.

*2.10.3. Testing of optimal wavelength combinations*

The performance of the wavelength-optimized ASEC method was tested in sample areas outside the training ROIs. Two scenarios were tested in this study. Firstly, for each given number of wavelengths ($n$ = 3, 4, 5, or 6), the corresponding optimal wavelength channels were extracted from the hyperspectral cube of the tested ROIs. Then, the procedure used to evaluate $\Delta E^*_{94}$ in the wavelength optimization step was followed: ASEC was applied to the extracted wavelength channels. The resulting estimated tristimulus values were compared against the ground truth to calculate the average $\Delta E^*_{94}$. This represents the expected color errors when our colorization technique is applied to images outside the training ROIs. The test results for $n$ = 3 are shown in the Results section.

Secondly, multiplexed multispectral lens-free imaging using three *simultaneous* illumination wavelengths ($n$ = 3) was conducted for the same sample. The three illumination wavelengths were the ones previously optimized in the training step. Image reconstruction was conducted using DPSR, TIE, and multi-height phase recovery, and ASEC was performed using the previously calculated $\mathbf{R_{aa}}$. The estimated color image was registered to the "ground truth" image. A pixelwise average $\Delta E^*_{94}$ was calculated between the estimated color image and the "ground truth" color image in the tested ROIs, which represents the expected color error of an independent imaging experiment of the sample using multiplexed illumination and reconstruction schemes.

### 2.11. Color-corrected lens-free holographic imaging procedure

The complete color-corrected wavelength-multiplexed lens-free imaging procedure for a new sample, whose tissue-type and stain combination is trained and optimized according to Section 2.10, is shown in **Figure 3**. First, wavelength-multiplexed illumination was used during the image acquisition to capture holograms with xyz translations. One of the low-resolution holograms on each lateral plane was wavelength-demultiplexed, and autofocusing was applied to find its z-distance. Then, DPSR was applied to the holograms measured on each lateral plane followed by phase retrieval using the pixel-super-resolved holograms. The phase-retrieved intensity images at multiple wavelengths were transformed to a color image with the ASEC method using the previously learned statistics of this specific tissue type and stain combination. The result is a holographic color image that exhibits an accurate color reproduction.

### 3. Results

We evaluated the ASEC performance using the testing methodology in Section 2.10. That is, the average $\Delta E^*_{94}$ value of each estimated color image with respect to the ground truth color image (i.e., from the lens-free hyperspectral imaging) was used to quantify the average color inaccuracy, owing to the capability of $\Delta E^*_{94}$ to account for the perceptual non-uniformity of widely-used color spaces [52]. However, please note that the average $\Delta E^*_{94}$ value will be partially influenced by the noise and artifacts within the image. The ASEC performance was compared to the direct

RGB combination approach in Figure 4 and **Figures S1–S3**. More specifically, the color performance of the lens-free color imaging approaches using sequential or multiplexed illumination, with or without ASEC, were compared qualitatively using visual judgement and quantitatively by calculating the average $\Delta E^*_{94}$. All three-wavelength imaging results (Figure 4 and Figures S1–S3, Columns (a)–(d)), except Figure 4 (c3), used the optimal three-wavelength combinations specific to each sample type (shown in **Figure 5**), which were identified from the wavelength optimization (training) procedure (Figure 2 (middle column) and Section 2.10). The only exception, Figure 4 (c3), used a suboptimal wavelength combination of 450 nm, 510 nm, and 580 nm, which is the best-performing (i.e., least-$\Delta E^*_{94}$) wavelength combination that is compatible with DPSR; the latter requires that the three wavelengths belong to the blue, green, and red spectral regions, respectively [10]. The optimal wavelength combination for "artery H&E" (510 nm, 560 nm, and 600 nm) does not satisfy this condition.

Evidently, the images using ASEC (Figure 4 and Figures S1–S3, Columns (b) and (d)) exhibit a better color accuracy than the corresponding images colorized with the direct RGB combination method (Columns (a) and (c)). Thus, they are in a much better agreement with the ground truth (Column (e)), both visually and quantitatively. This is reflected by the average $\Delta E^*_{94}$ values displayed in the bottom right of each individual image in Columns (a)–(d). Regarding the 25 tested samples, the average $\Delta E^*_{94}$ is improved (reduced) by approximately 2.48 and 1.73 after applying ASEC (compared to the direct RGB combination method) to approximately 1.71 and 3.83 for the sequential and multiplexed illumination cases, respectively.

The $\Delta E^*_{94}$ values of the simultaneous (i.e., multiplexed) illumination results using ASEC (Figure 4 and Figures S1–S3, Column (d)) are, on average, larger than those with sequential illumination, by approximately 2.12 (Column (b)), which is mainly due to the additional steps involved in DPSR inevitably introducing additional error. However, for 22 of 25 samples, the $\Delta E^*_{94}$ of the simultaneous illumination results with ASEC is below 5, which demonstrates a decent color reproduction.

To further reduce the data amount required for color imaging on the basis of wavelength-multiplexed imaging and ASEC, we applied a sparsity constraint for performing multi-height phase recovery [27] and reduced the number of hologram heights from 8 to 4 and 2. The results are shown in Figure 5. With the use of the sparsity constraint, the ASEC-colorized images using four heights (Column (c)) still have a decent quality, whereas those reconstructed from two heights show relatively severe reconstruction artifacts.

## 4. Discussion

In the ASEC method, we chose to develop our linear estimator based on the absorbance because in our transmission mode imaging system, the spectral absorbance of the sample and the amount of dye or pigment at each lateral (*x*, *y*) position within the sample follow a linear relation given by the Beer–Lambert law:

$$A(\lambda) = \sum_{j=1}^{n_{\text{species}}} A_j(\lambda) = \sum_{j=1}^{n_{\text{species}}} \varepsilon_j(\lambda) \int_0^l c_j(z) \, dz \qquad (13)$$

where $A(\lambda)$ is the absorbance of the sample at wavelength $\lambda$, $\varepsilon_j(\lambda)$ the molar attenuation coefficient or absorptivity of the attenuating species *j* in the sample, $n_{\text{species}}$ refers to the total number of attenuating species within the sample, $c_j$ the molar concentration of the attenuating species *j*, and *l* the optical path length of the illumination light going through the sample.

The optimal three-wavelength combinations of all the samples are shown in **Figure 6**. The average wavelengths are 450.0 nm, 535.2 nm, and 605.6 nm, respectively, and represented by solid lines. Interestingly, the majority of the optimal wavelengths are clustered around these three values. Although we think that this is partially related to our wavelength optimization procedure that aims for accurate human color perception (whose peak sensitivities are approximately 446 nm, 555 nm, and 599 nm) [53], the optimal wavelength combinations are not entirely intuitive. This is especially evident in the selection of the green wavelength: its average value (535.2 nm) is approximately 20 nm shorter than the sensitivity peak of an average human (approximately 555 nm). Moreover, a distinct outlier is presented by the artery tissue with H&E stains (dashed gray box in **Figure 6**). Its optimized wavelengths are 510 nm, 560 nm, and 600 nm, respectively. This sample possesses larger areas with a very high absorbance for 520–540 nm (**Figure S4**), which may cause the optimization to favor the wavelength combinations with denser sampling in the green region.

The ranked wavelength combinations obtained in the training process (Section 2.10) for all 25 samples are shown in the **Tables S1–S25**, thereby providing a practical guide to wavelength selection for holographic imaging of various pathology slides, involving different dyes and tissue types. Due to space constraints, the top five combinations for each number of wavelengths $n$ ($n = 3, 4, 5, 6$) are displayed. According to these tables, the usage of the optimal wavelength combination for a larger $n$ always leads to more color-accurate results (i.e., lower average $\Delta E^*_{94}$). Indeed, even the fifth-best wavelength combination of a larger $n$ always results in more accurate colorization than the absolute best of a smaller $n$.

However, the improved color accuracy offered by using more wavelengths can still be nullified if those wavelengths are not chosen carefully. **Figure 7** shows the $\Delta E^*_{94}$ averaged across all the samples and plotted as a function of the relative ranking of each wavelength combination obtained in the training process (Section 2.10), starting from the lowest $\Delta E^*_{94}$ to the highest. In other words, the $x$-axis presents the rank of each wavelength combination (in an ascending order) normalized by the total number of possible combinations for a given $n$. The error bars represent the standard error of the mean (SEM) of the averaged $\Delta E^*_{94}$ values of the 25 samples. According to **Figure 7**, for example the four-wavelength combination ranking at the top 2.65 %, the five-wavelength combination ranking at the top 12.5 %, and the six-wavelength combination ranking at the top 28.7 % result in an equivalent average $\Delta E^*_{94}$ value as the best three-wavelength combination. Therefore, a poor wavelength selection will defeat the purpose to improve color accuracy with denser sampling in the optical spectrum. Moreover, even if an optimal wavelength selection is assumed, the improvement in $\Delta E^*_{94}$ by adding extra wavelengths beyond three is modest. For an average sample, the reduction in $\Delta E^*_{94}$ is 0.38 when going from three to four, 0.27 from four to five, and 0.21 from five to six wavelengths, which would be barely noticeable to the eye.

## 5. Conclusion

An ASEC method for accurate-color holographic imaging of pathology slides was presented. The method converts the holographic reconstruction images of the sample at a small number of wavelengths ($n = 3, 4, 5,$ or $6$) to an image with accurate color reproduction by computationally estimating its absorbance spectrum at each pixel position. In addition, we extracted the spectral statistics of 25 samples with different tissue–stain combinations and performed global optimization of the $n$-wavelength selection ($n = 3, 4, 5, 6$) for each sample, respectively, thereby demonstrating a significant improvement in the color accuracy compared to the direct RGB combination method. Furthermore, ASEC can be readily combined with multiplexed-wavelength holographic imaging using DPSR, leading to accurate color reproduction with a significantly reduced number of raw measurements. In conclusion, this study provides a practical guide and a database to achieve accurate-color holographic imaging of pathology slides, spanning different dyes and tissue types.

**Acknowledgements**

The Ozcan Research Group at UCLA acknowledges the support of NSF Engineering Research Center (ERC, PATHS-UP), the Army Research Office (ARO; W911NF-13-1-0419 and W911NF-13-1-0197), the ARO Life Sciences Division, the National Science Foundation (NSF) CBET Division Biophotonics Program, the NSF Emerging Frontiers in Research and Innovation (EFRI) Award, the NSF INSPIRE Award, NSF Partnerships for Innovation: Building Innovation Capacity (PFI:BIC) Program, the National Institutes of Health (NIH, R21EB023115), the Howard Hughes Medical Institute (HHMI), Vodafone Americas Foundation, the Mary Kay Foundation, and Steven & Alexandra Cohen Foundation. The authors also acknowledge the Histology Lab and the Translational Pathology Core Laboratory (TPCL) at UCLA for preparing the pathology slides; and Dr. Jonathan Zuckerman from the Pathology Department at UCLA for his help with selecting the pathology samples.

**Figures and Tables**

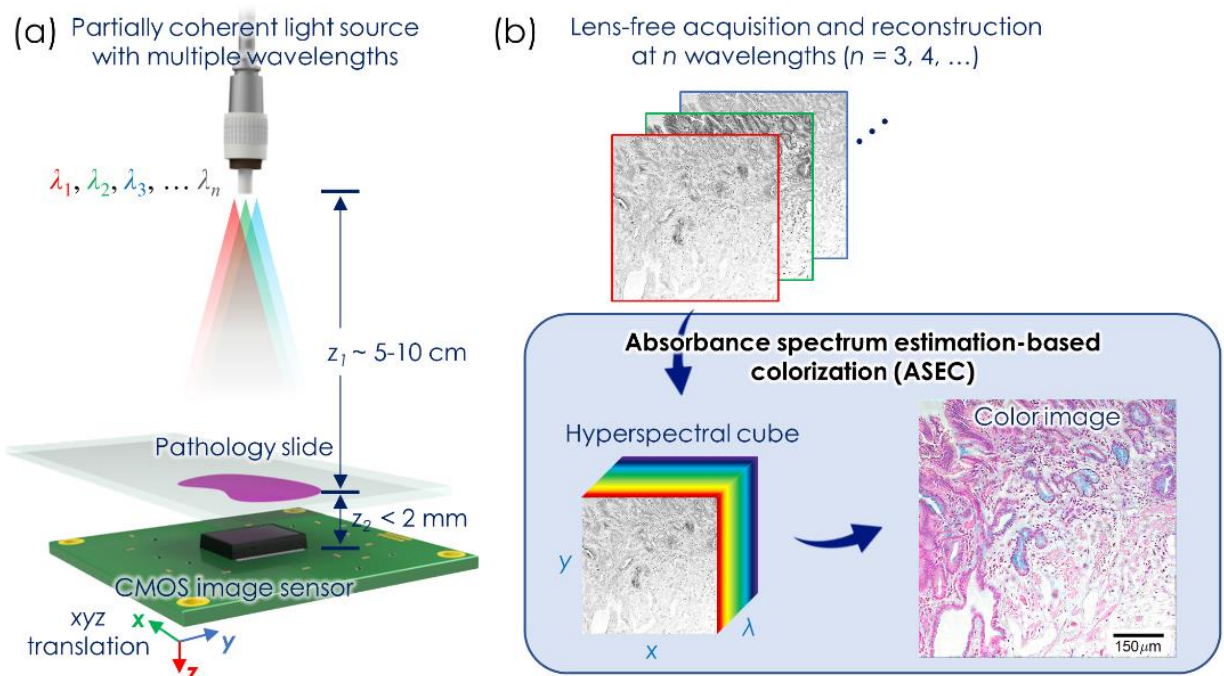

**Figure 1**. Schematic of accurate-color lens-free holographic microscopy method. (a) Schematic of the optical setup. A partially coherent light source with multiple wavelengths illuminates the pathology slide simultaneously or sequentially (depending on the mode of operation). An image sensor is placed underneath the sample slide at a short axial distance ($z_2 < 2$ mm) to capture in-line holograms of the sample. (b) After the lens-free image acquisition and reconstruction, an absorbance spectrum estimation-based colorization (ASEC) method is applied, resulting in a color-accurate image of the sample.

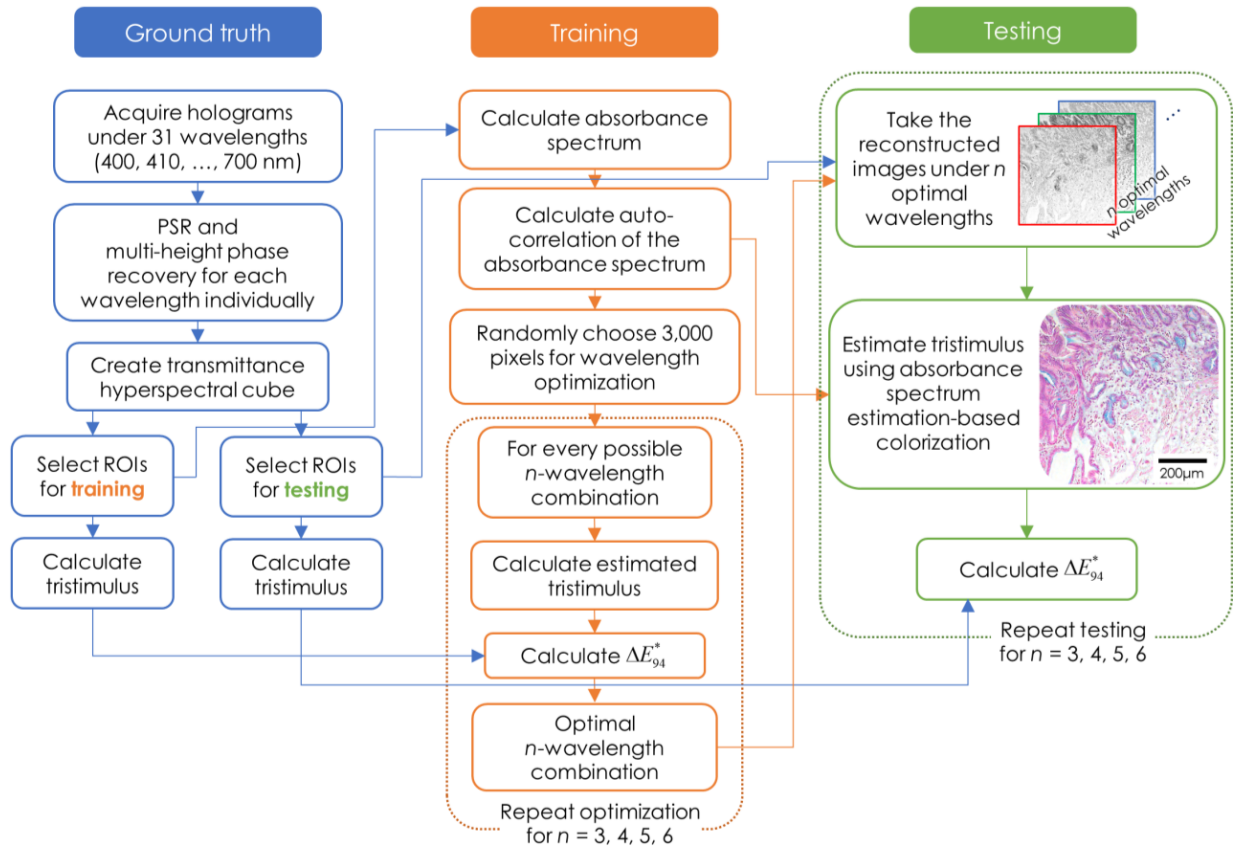

**Figure 2.** Collection of the spectral statistics and wavelength optimization (training) as well as quantification of the average color error (testing) using the ASEC method.

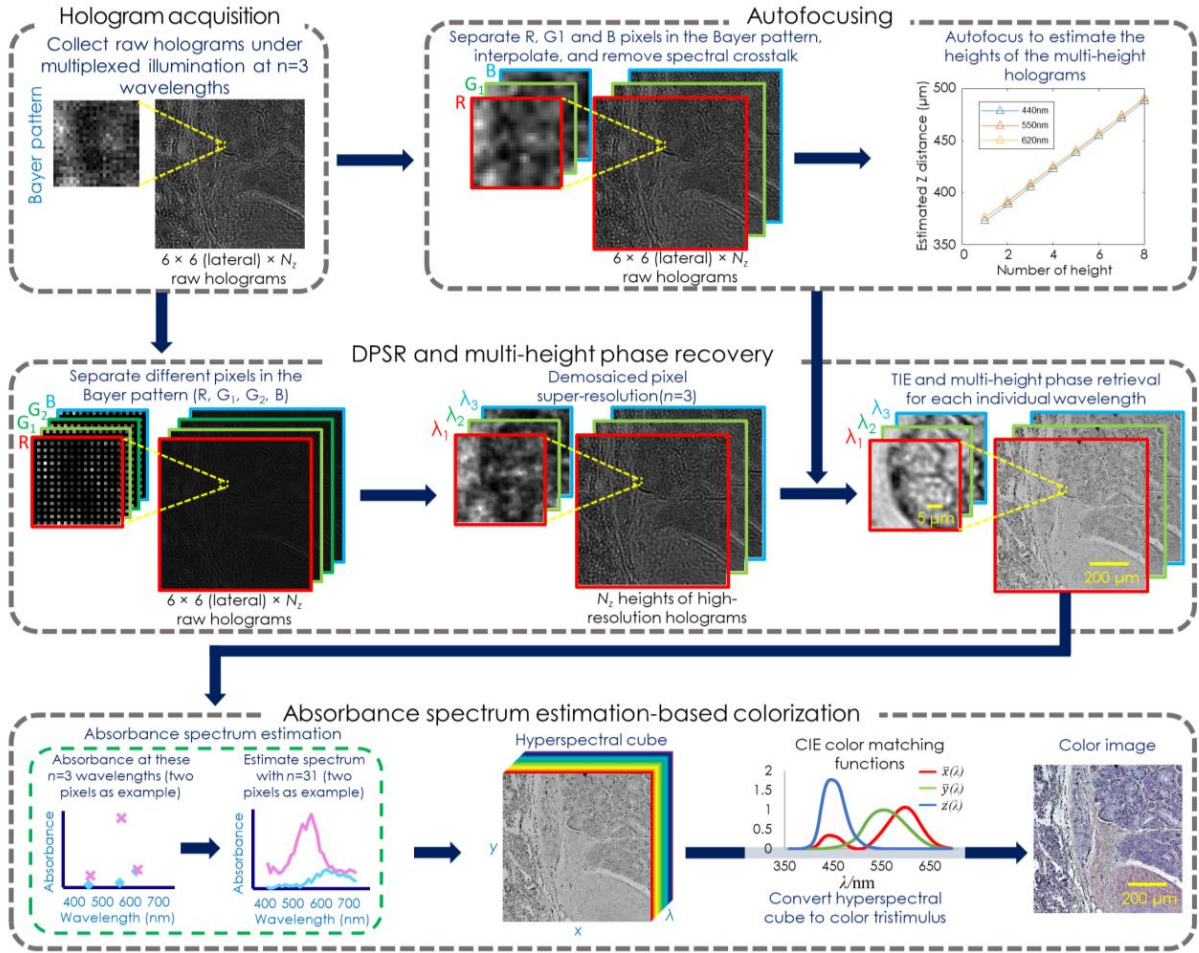

**Figure 3.** Color-corrected and wavelength-multiplexed lens-free holographic imaging procedure. $N_z$ is the number of the hologram measurement heights.

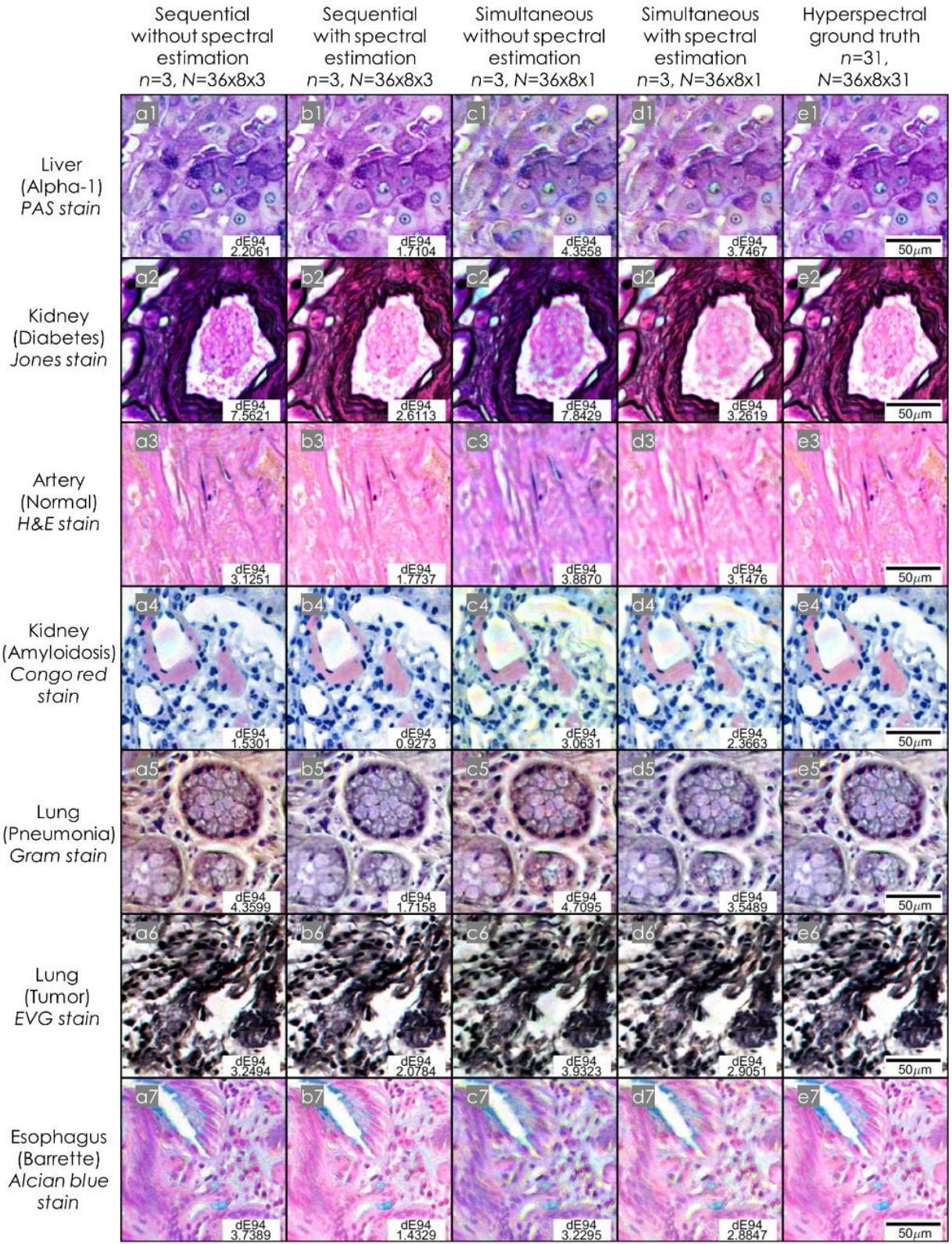

**Figure 4.** Comparison of the performance of various lens-free imaging and colorization schemes for different samples. To the left of each row, the tissue type, its pathological condition, and its stain are specified. **Column (a):** The holographic reconstruction intensity images at the three

optimal wavelengths using *sequential* illumination are directly placed into the corresponding R, G, and B channels to form an RGB image. **Column (b):** ASEC is performed on the same data as in Column (a). **Column (c):** The holographic reconstruction intensity images at the three optimal wavelengths (except c3, which uses a suboptimal combination that is compatible with DPSR) using *simultaneous* illumination and wavelength demultiplexing are directly placed into the corresponding R, G, and B channels to form an RGB image. **Column (d):** ASEC is performed on the same data as in Column (c). **Column (e):** The ground truth color images generated from hyperspectral lens-free imaging with 31 wavelengths, ranging from 400 nm to 700 nm. The "dE94" corresponds to their respective $\Delta E^*_{94}$ value with respect to the ground truth color images in Column (e); $n$ is the number of wavelengths; $N$ is the number of holograms used to reconstruct the images corresponding to each column and equal to the multiplication of the number of lateral shifts for PSR or DPSR, with the number of heights, and the number of measurements per shifting position for multispectral imaging.

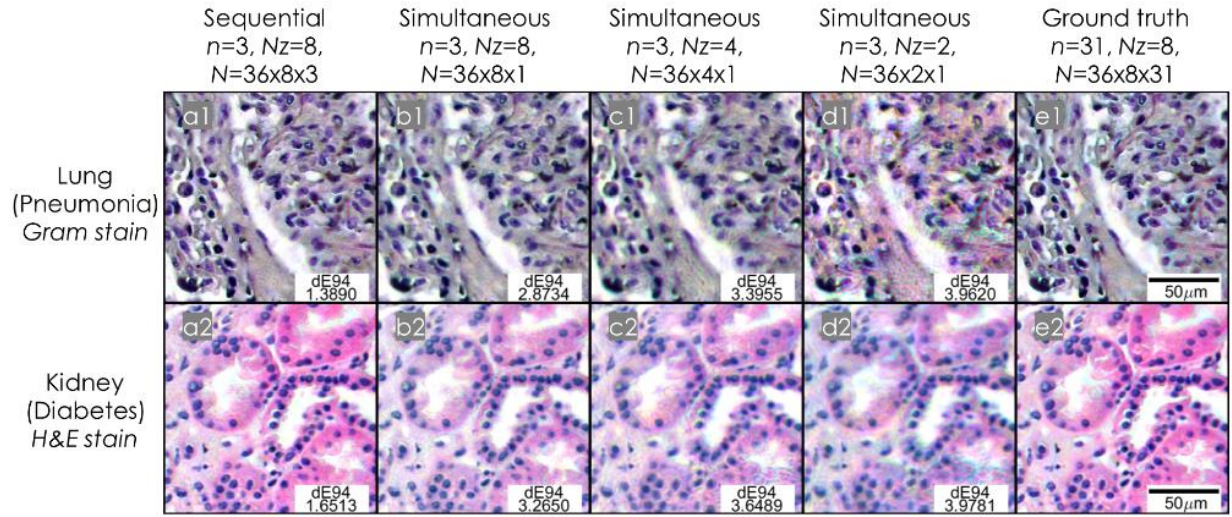

**Figure 5.** Reducing the number of hologram heights using multi-height phase recovery with sparsity constraint. $N_z$ is the number of the hologram measurement heights; $N$ is the total number of acquired holograms used for reconstruction.

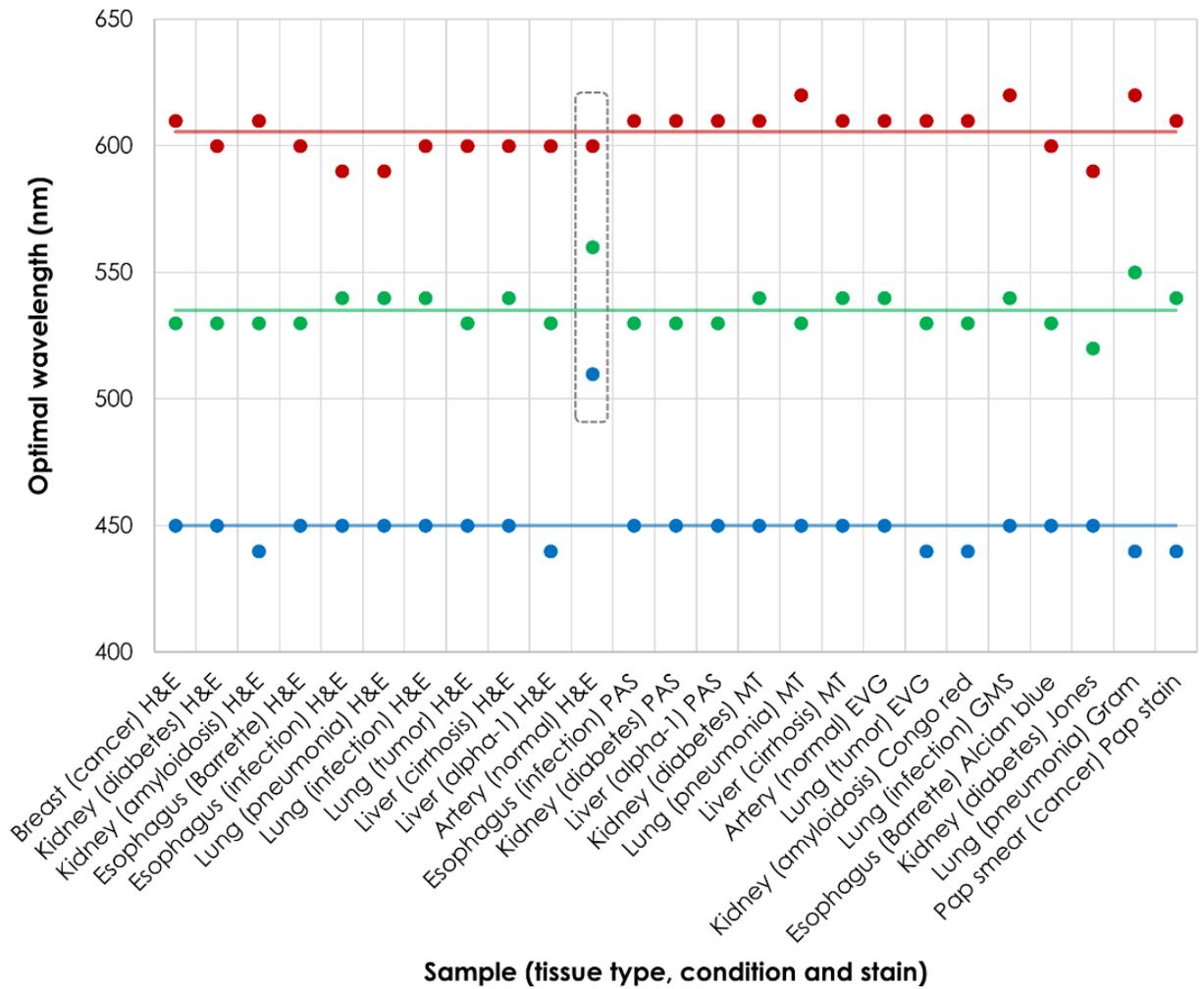

**Figure 6.** Optimal three-wavelength combinations for various pathology samples. The solid lines represent the averaged optimal wavelengths for all the samples. The dashed gray box corresponds to an outlier: artery with H&E stain. Its optimal wavelengths deviate far from the average.

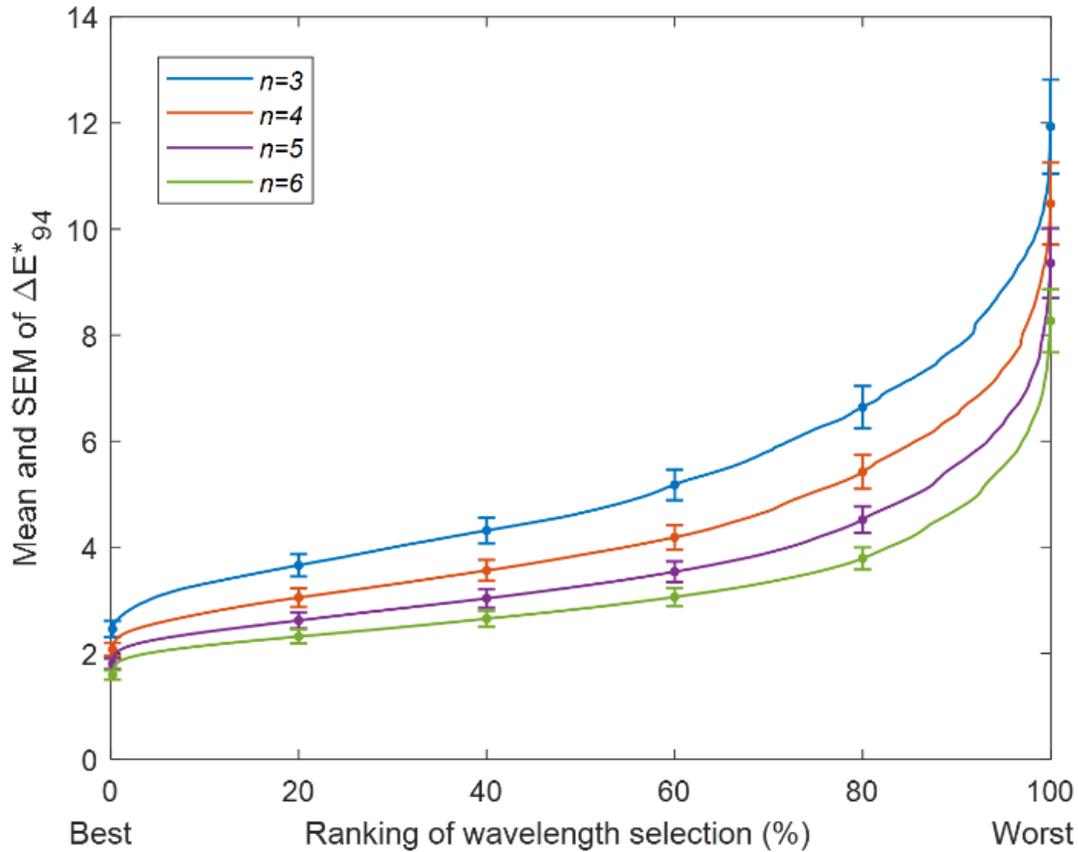

**Figure 7.** Average ΔE*$_{94}$ for 25 different samples for different wavelength combinations. The x-axis presents the relative ranking for each given wavelength combination with respect to all possible $n$-wavelength combinations ($n$ = 3, 4, 5, 6). Smaller relative rankings correspond to better color accuracy, i.e., lower average ΔE*$_{94}$. Error bars are shown at six equally-spaced points, representing the standard error of the mean (SEM) of the average ΔE*$_{94}$ values regarding the 25 samples.